# Randomized Mixture Models for Probability Density Approximation and Estimation

Hien D. Nguyen[*,1], Dianhui Wang[2,3], and Geoffrey J. McLachlan[4]

April 22, 2018

**Abstract**

Randomized neural networks (NNs) are an interesting alternative to conventional NNs that are more used for data modeling. The random vector functional-link (RVFL) network is an established and theoretically well-grounded randomized learning model. A key theoretical result for RVFL networks is that they provide universal approximation for continuous maps, on average, almost surely. We specialize and modify this result, and show that RFVL networks can provide functional approximations that converge in Kullback-Leibler divergence, when the target function is a probability density function. Expanding on the approximation results, we demonstrate the the RFVL networks lead to a simple randomized mixture model (MM) construction for density estimation from random data. An expectation-maximization (EM) algorithm is derived for the maximum likelihood estimation of our randomized MM. The EM algorithm is proved to be globally convergent and the maximum likelihood estimator is proved to be consistent. A set of simulation studies is given to provide empirical evidence towards our approximation and density estimation results.

---
[*]Corresponding Author (Email: h.nguyen5@latrobe.edu.au). [1]Department of Mathematics and Statistics, La Trobe University, Bundoora 3086, Melbourne Victoria, Australia. [2]Department of Computer Science and Information Technology, La Trobe University, Bundoora 3086, Melbourne Victoria, Australia. [3]State Key Laboratory of Synthetical Automation for Process Industries, Northeastern University, Shenyang 110819, China. [4]School of Mathematics and Physics, University of Queensland, St. Lucia 4072, Brisbane Queensland, Australia.

# 1 Introduction

Neural networks (NNs) have become a ubiquitous tool for predictive modeling and data analytics across numerous application domains. As noted in the introduction of [12], the usefulness of NNs can be attributed to their numerous beneficial attributes such as their learning and representational capabilities, when applied to the analysis of nonlinear data and signals.

Randomized NNs have become an increasingly popular topic of modern research, and there are now many available randomized NNs in the literature. Such networks include the randomized radial basis functions networks [2], the random vector functional-link networks [14], and the stochastic configuration networks [31]. A recent review of the randomized NNs literature can be found in [27].

Mixture models (MMs; [22]) can be viewed as NNs that are adapted for functional approximation in the spaces of probability distributions or density functions (see for example [26], and [33]). There is a folk theorem which states that the class of MM densities can approximate any probability density function (PDF) if the number of mixture components of the MM is taken to be sufficiently large. Examples of such statements include "any continuous distribution can be approximated arbitrarily well by a finite mixture of normal densities with common variance (or covariance matrix in the multivariate case)" [22, p. 176] and "provided the number of component densities is not bounded above, certain forms of mixture can be used to provide arbitrarily close approximations to a given probability distribution" [28, p. 50].

We note that MM forms have been explored in the randomized NN literature, such as in [13] and [30]. In the former publication, the author utilizes MMs in order to model probabilistic uncertainty around an RVFL mean regression model. In the latter publication, the authors utilize a kernel density estimator (a kind of non-parametric MM) in order to determine penalty weights for observations that are used to train an SCN. Both of these contributions are different to our contributions in this article.

Let $\mathcal{D}(\mathbb{X}) = \left\{f \in \mathcal{C}(\mathbb{X}) : f \geq 0, \int_{\mathbb{X}} f d\boldsymbol{x} = 1\right\}$ be the class of valid PDFs, where $\mathbb{X} \subset \mathbb{R}^d$ be compact for some $d \in \mathbb{N}$ (zero exclusive) and $\mathcal{C}(\mathbb{X})$ be the class of continuous functions with support $\mathbb{X}$. Define a function $f$ to be in the class $\mathcal{L}_p(\mathbb{X})$ if $\|f\|_{p,\mathbb{X}} = \left(\int_{\mathbb{X}} |f|^p d\boldsymbol{x}\right)^{1/p} < \infty$, where we refer to $\|\cdot\|_{p,\mathbb{X}}$ as the $\mathcal{L}_p(\mathbb{X})$ norm. Furthermore, say that a function $\phi$ is a "hump-shaped" marginal PDF (cf. [20, Sec. 9.3.4]) if $\phi \in \mathcal{H}(\mathbb{R})$, where $\mathcal{H}(\mathbb{R}) = \left\{\phi \in \mathcal{D}(\mathbb{R}) : \phi(x) = \psi(x^2), \psi \text{ is decreasing on } \mathbb{R}_+\right\}$ and $\mathbb{R}_+ = (0, \infty)$. Let $\boldsymbol{x}^\top = (x_1, \ldots, x_d) \in \mathbb{R}^d$ and $\boldsymbol{y}^\top = (y_1, \ldots, y_d) \in \mathbb{R}^d$, where $(\cdot)^\top$ is the transposition operator. The following version of [5, Thm. 33.1] provides a concrete form of the folk theorem (see also [23]).

**Theorem 1.** *Let $f$ be a PDF in $\mathcal{D}(\mathbb{R}^d)$ and let $\phi$ be a PDF in $\mathcal{H}(\mathbb{R})$. Define the class of location-scale mixtures of $\phi$ to be*

$$\mathcal{M}_\phi = \left\{m : \; m(\boldsymbol{x}) = \int_{\mathbb{R}^+} \int_{\mathbb{R}^d} w^d \prod_{j=1}^d \phi(wx_j - wy_j) p_{\boldsymbol{Y}}(\boldsymbol{y}) p_W(w) \, d\boldsymbol{y} \, dw\right\},$$

*where $p_{\boldsymbol{Y}}$ and $p_W$ are PDFs on the supports $\mathbb{R}^d$ and $\mathbb{R}_+$, respectively. If $\epsilon > 0$*

*and $p \in [1, \infty]$, then there exists an $m \in \mathcal{M}_\phi$ such that $\|f - m\|_{p,\mathbb{X}} < \epsilon$, for any compact $\mathbb{X} \subset \mathbb{R}^d$.*

Using theoretical tools from functional analysis, and from [14] and [33], we provide a version of Theorem 1 for randomized MMs. That is, we prove that a mixture of $n$ randomly sampled hump-shaped PDFs from some probability measure can approximate any PDF in the $\mathcal{C}(\mathbb{X})$ norm, under mild regularity conditions as $n$ approaches infinity. The precise nature of our theoretical results will be elucidated in the sequel. Furthermore, we demonstrate that control of the $\mathcal{L}_2(\mathbb{X})$ norm also allows for control of the Kullback-Leibler (KL) divergence [16].

Apart from our approximation results, we also demonstrate how our derived randomized MMs can be applied to density estimation from sampled data from an unknown population. We construct density estimators using a modified EM algorithm (see [6] and [21]) for maximum likelihood estimation given some randomly sampled mixture components. A guarantee of large-sample performance is provided via a consistency result.

To conclude our article, we perform a pair of short simulation studies. The first simulation study seeks to demonstrate the function approximation capacity of our randomized MMs. The second simulation demonstrates the density estimation capabilities of our randomized MMs when fitted to sampled data.

The remainder of the article proceeds as follows. The main approximation results are presented in Section 2. Density estimation results are presented in Section 3. Simulations are performed in Section 4. Conclusions are drawn in Section 5.

## 2  Main Results

Retain the notation from Section 1 and let $\mathbb{I} = [0,1]^d$ be the unit hypercube of dimension $d$. Further define $\boldsymbol{w}^\top = (w_1, \ldots, w_d) \in \mathbb{R}_+^d$ and utilize the convention of lowercase letters for realizations and uppercase letters for random variables. For $n \in \mathbb{N}$, let $\boldsymbol{Z}_n^\top = (\boldsymbol{W}_1^\top, \ldots, \boldsymbol{W}_n^\top, \boldsymbol{Y}_1^\top, \ldots, \boldsymbol{Y}_n^\top)$ and let $A_i = a(\boldsymbol{Y}_i)$ for each $i \in [n] = \{1, \ldots, n\}$, where $a(\boldsymbol{y})$ is a real-valued function. Define the distance $\rho_\mathbb{X}(f, g)$ between two potentially stochastic functions $f$ and $g$ via the equation

$$\rho_\mathbb{X}^2(f, g) = \mathbb{E} \int_\mathbb{X} [f(\boldsymbol{x}) - g(\boldsymbol{x})]^2 \, d\boldsymbol{x},$$

where the expectation (i.e. $\mathbb{E}$) is taken over the data generating process of $f$ and $g$. Suppose that $f \in \mathcal{C}(\mathbb{I}^d)$ and approximate $f$ via the random mixture

$$f(\boldsymbol{x}; \boldsymbol{Z}_n) = \frac{1}{n} \sum_{i=1}^n A_i \prod_{j=1}^d W_{ij} \phi(W_{ij} x_j - W_{ij} Y_{ij}). \tag{1}$$

The following result can be inferred from Theorem 2 of [14] and establishes the capacity of (1) as an approximator for $f$.

**Theorem 2.** *If $\mathbb{X} \subsetneq \mathbb{I}^d$ is compact, $f \in \mathcal{C}(\mathbb{I}^d)$, and $\phi \in \mathcal{D}(\mathbb{I}^d)$, then: (a) Uniformly over $\mathbb{X}$, we have*

$$f(\boldsymbol{x}) = \lim_{w_1 \to \infty} \cdots \lim_{w_d \to \infty} \int_{\mathbb{I}^d} g(\boldsymbol{x}; \boldsymbol{w}, \boldsymbol{y}) \, d\boldsymbol{y}$$

*and*

$$\lim_{w_1 \to \infty} \cdots \lim_{w_d \to \infty} \int_{\mathbb{I}^d} g(\boldsymbol{x}; \boldsymbol{w}, \boldsymbol{y}) \, d\boldsymbol{y} = \lim_{\omega \to \infty} \omega^{-d} \int_{\mathbb{I}^d \times \Omega^d} g(\boldsymbol{x}; \boldsymbol{w}, \boldsymbol{y}) \, d\boldsymbol{y} \, d\boldsymbol{w},$$

*where $\Omega = [0, \omega]$ and $g(\boldsymbol{x}; \boldsymbol{w}, \boldsymbol{y}) = \prod_{j=1}^d w_j \phi(w_j x_j - w_j y_j) f(\boldsymbol{y})$. (b) For any fixed $\omega \in \mathbb{R}_+$, there exists data generating processes for $\boldsymbol{W}$ and $\boldsymbol{Y}$ and a real-* *valued function* $a(\boldsymbol{y})$, *such that*

$$\rho_{\mathbb{X}}\left(f(\boldsymbol{x};\boldsymbol{Z}_n), \frac{1}{\omega^d}\int_{\mathbb{I}^d\times\Omega^d} g(\boldsymbol{x};\boldsymbol{w},\boldsymbol{y})\,d\boldsymbol{y}\,d\boldsymbol{w}\right) \to 0,$$

*as* $n \to \infty$.

A deficiency of Theorem 2 is the restriction to the unit hypercube domain, which is unnatural in many density approximation and estimation problems. The following result resolves this deficiency and is better suited for the problem of density approximation.

**Theorem 3.** *If $\mathbb{X} \subset \mathbb{R}^d$ is compact, $f \in \mathcal{C}(\mathbb{R}^d)$, and $\phi \in \mathcal{D}(\mathbb{R}^d)$, then: (a) Uniformly over $\mathbb{X}$, we have*

$$f(\boldsymbol{x}) = \lim_{w_1 \to \infty} \cdots \lim_{w_d \to \infty} \int_{\mathbb{R}^d} g(\boldsymbol{x};\boldsymbol{w},\boldsymbol{y})\,d\boldsymbol{y} \qquad (2)$$

*and*

$$\lim_{w_1 \to \infty} \cdots \lim_{w_d \to \infty} \int_{\mathbb{R}^d} g(\boldsymbol{x};\boldsymbol{w},\boldsymbol{y})\,d\boldsymbol{y} = \lim_{\omega \to \infty} \omega^{-d} \int_{\mathbb{R}^d \times \Omega^d} g(\boldsymbol{x};\boldsymbol{w},\boldsymbol{y})\,d\boldsymbol{y}\,d\boldsymbol{w}, \qquad (3)$$

*where $\Omega = [0,\omega]$ and $g(\boldsymbol{x};\boldsymbol{w},\boldsymbol{y}) = \prod_{j=1}^d w_j \phi(w_j x_j - w_j y_j) f(\boldsymbol{y})$. (b) For any fixed $\omega \in \mathbb{R}_+$ and compact set $\mathbb{K} \supset \mathbb{X}$, there exists data generating processes for $\boldsymbol{W}$ and $\boldsymbol{Y}$ and a real-valued function $a(\boldsymbol{y})$, such that*

$$\rho_{\mathbb{X}}\left(f(\boldsymbol{x};\boldsymbol{Z}_n), \frac{1}{\omega^d}\int_{\mathbb{K}\times\Omega^d} g(\boldsymbol{x};\boldsymbol{w},\boldsymbol{y})\,d\boldsymbol{y}\,d\boldsymbol{w}\right) \to 0, \qquad (4)$$

*as* $n \to \infty$.

The proof of Theorem 3 follows in the same vein as that of Theorem 2. For completeness, we provide the proof of Theorem 3, below. Define $\|\cdot\|$ to be the usual Euclidean norm and $\star$ to be the convolution operator. The following lemma appears in [3, Ch. 20].

**Lemma 1.** *Let $h_1, h_2, \ldots$ be a sequence of functions in $\mathcal{L}_1(\mathbb{R}^d)$ such that (a) $\sup_w \|h_w\|_1 < \infty$ and (b) $\int_{\mathbb{R}^d} h_w(\boldsymbol{x}) = 1$, for all $w \in \mathbb{N}$. If we further assume that (c) $\lim_{w \to \infty} \int_{\|\boldsymbol{x}\| > \delta} |h_w(\boldsymbol{x})| = 0$ for every $\delta \in \mathbb{R}_+$, then for each bounded function $f \in \mathcal{C}(\mathbb{R}^d)$, we have $h_w \star f \to f$ as $w \to \infty$, uniformly over every compact subset of $\mathbb{R}^d$.*

Let $h_w(\boldsymbol{x}) = \prod_{j=1}^{d} w_j(w) \phi(w_j(w) x_j)$, where $w_j(w)$ is an increasing sequence in $w \in \mathbb{N}$, $w_j(w) \in \mathbb{R}_+$ for each $w$, and $w_j(w) \to \infty$ for each $j \in [d]$. We seek to show that the sequence $\{h_w\}$ satisfies assumptions (a)–(c) of Lemma 1.

Firstly, note that $\phi \in \mathcal{H}(\mathbb{R})$, which implies that $\phi \in \mathcal{D}(\mathbb{R})$. This implies that $\phi$ is the marginal PDF of some random variable $X \in \mathbb{R}$. Using the change-of-variable formula (cf. [1, Thm. 3.6.1]), we know that $w\phi(wx) \in \mathcal{D}(\mathbb{R})$ is also a marginal PDF of some random variable $X$, for any $w \in \mathbb{R}_+$. Thus, $h_w \in \mathcal{D}(\mathbb{R}^d)$ is a joint PDF of independent random variables $X_j$, each with marginal PDFs $w_j(w) \phi(w_j(w) x_j)$, for $j \in [d]$. We have therefore validated Assumptions (a) and (b). Write $w_j = w_j(w)$ for each $j \in [d]$ and by Tonelli's theorem

$$\begin{aligned}
\int_{\|\boldsymbol{x}\|<\delta} h_w(\boldsymbol{x}) \, \mathrm{d}\boldsymbol{x} &\leq \int_{|x_1|<\delta} \cdots \int_{|x_d|<\delta} \prod_{j=1}^{d} w_j \phi(w_j x_j) \, \mathrm{d}\boldsymbol{x} \\
&= \prod_{j=1}^{d} \int_{|\xi_j|<w_j \delta} \phi(\xi_j) \, \mathrm{d}\xi_j \\
&\to 0,
\end{aligned}$$

as $w \to 0$, since $w_j$ is increasing and the tail of $\phi \in \mathcal{H}(\mathbb{R})$ is decreasing in volume. We have therefore validated (c) and can state the following result.

**Lemma 2.** *If $h_w(\boldsymbol{x}) = \prod_{j=1}^{d} w_j(w) \phi(w_j(w) x_j)$, where $w_j(w)$ is an increasing sequence in $w \in \mathbb{N}$, $w_j(w) \in \mathbb{R}_+$ for each $w$, and $w_j(w) \to \infty$ for each $j \in [d]$, then the sequence $\{h_w\}$ satisfies the assumptions and permits the conclusion of*

*Lemma 1.*

By the definition of a convolution, we have Equation 2 from Theorem 3. Repeated applications of l'Hôpital's rule, as per [14, Eqn. 27], yields Equation 3.

Fix $\omega \in \mathbb{R}_+$ and let $\boldsymbol{W}_1, \ldots, \boldsymbol{W}_n$ and $\boldsymbol{Y}_1, \ldots, \boldsymbol{Y}_n$ be two IID (independent and identically distributed) samples from data generating processes that are characterized by the density functions $p_{\boldsymbol{W}}$ and $p_{\boldsymbol{Y}}$, respectively. Let $U_1, \ldots U_n$ be an IID random sample with finite first and second moments, and define $\bar{U}_n = n^{-1} \sum_{i=1}^n U_i$ to be the sample mean. By the definition of the variance, we have $\mathbb{E} \left( \bar{U}_n - \mathbb{E}\bar{U}_n \right)^2 = n^{-1} \left( \mathbb{E}U_1^2 - [\mathbb{E}U_1]^2 \right)$. Notice that $\mathbb{E}\bar{U}_n = \mathbb{E}U_1$.

Write the left-hand side (LHS) of (4) as

$$\mathbb{E} \int_{\mathbb{X}} \left[ f(\boldsymbol{x}; \boldsymbol{Z}_n) - \frac{1}{\omega^d} \int_{\mathbb{K} \times \Omega^d} g(\boldsymbol{x}; \boldsymbol{w}, \boldsymbol{y}) \, \mathrm{d}\boldsymbol{y} \mathrm{d}\boldsymbol{w} \right]^2 \mathrm{d}\boldsymbol{x},$$

which equals

$$\int_{\mathbb{X}} \mathbb{E} \left[ f(\boldsymbol{x}; \boldsymbol{Z}_n) - \frac{1}{\omega^d} \int_{\mathbb{K} \times \Omega^d} g(\boldsymbol{x}; \boldsymbol{w}, \boldsymbol{y}) \, \mathrm{d}\boldsymbol{y} \mathrm{d}\boldsymbol{w} \right]^2 \mathrm{d}\boldsymbol{x}$$

by Fubini's theorem. Define $U_i = A_i \prod_{j=1}^d W_{ij} \phi(W_{ij} x_j - W_{ij} Y_{ij})$ for each $i \in [n]$ and suppose that $\mathbb{E}U_1 = \omega^{-d} \int_{\mathbb{K} \times \Omega^d} g(\boldsymbol{x}; \boldsymbol{w}, \boldsymbol{y}) \, \mathrm{d}\boldsymbol{y} \mathrm{d}\boldsymbol{w}$. We require an appropriate function $a$, and densities $p_{\boldsymbol{W}}$ and $p_{\boldsymbol{Y}}$, in order to ensure compatibility of the aforementioned definitions. Take $p_{\boldsymbol{W}}(\boldsymbol{w}) = \omega^{-d}$ to be the uniform distribution over the hypercube $\Omega^d$ and select any $p_{\boldsymbol{Y}} \in \mathcal{D}(\mathbb{K})$ that is bounded away from zero. Further, let $a(\boldsymbol{y}) = f(\boldsymbol{y})/p_{\boldsymbol{Y}}(\boldsymbol{y})$. We can check that our definitions are now compatible, and thus we can write

$$\rho_{\mathbb{X}}^2 \left( f(\boldsymbol{x}; \boldsymbol{Z}_n), \frac{1}{\omega^d} \int_{\mathbb{K} \times \Omega^d} g(\boldsymbol{x}; \boldsymbol{w}, \boldsymbol{y}) \, \mathrm{d}\boldsymbol{y} \mathrm{d}\boldsymbol{w} \right) = \frac{1}{n} \int_{\mathbb{X}} \left( \mathbb{E}U_1^2 - [\mathbb{E}U_1]^2 \right).$$

Finally, it suffices to note that $\mathbb{E}U_1^2 - (\mathbb{E}U_1)^2$ exists and independent of $n$ by the assumptions on the spaces from which $f$, $\phi$, and $p_{\boldsymbol{Y}}$ are drawn. The desired conclusion can be obtained by taking the limit as $n \to \infty$.

We note that Theorem 2 can be obtained from Theorem 3 by setting $\mathbb{X} \subsetneq \mathbb{I}^d$ and $\mathbb{K} = \mathbb{I}^d$. Furthermore, we draw the following interesting corollaries from Theorem 3.

**Corollary 1.** *If we assume that $f \in \mathcal{D}(\mathbb{R}^d)$ in addition to the conditions of Theorem 3, then: (a) For any fixed $\boldsymbol{z}_n$ and $\boldsymbol{x} \in \mathbb{X}$, $f(\boldsymbol{x}; \boldsymbol{z}_n) \geq 0$. (b) If $c = \int_{\mathbb{K}} f \, d\boldsymbol{x}$ and $p_{\boldsymbol{Y}} = c^{-1} f$, then we can set $a(\boldsymbol{y}) = c$ in order satisfy (4). (c) If we let $\mathbb{K}$ be a hypercube that with sides of length $\beta$, and we let $p_{\boldsymbol{Y}}(\boldsymbol{y}) = \beta^{-d}$ be the uniform distribution over $\mathbb{K}$, then we can set $a(\boldsymbol{y}) = \beta^d f(\boldsymbol{y})$ in order to satisfy (4).*

*Proof.* Part (a) can be proved by observing that $a = f/p$ must always be positive, as are $w$ and $\phi$. Thus their products and positive sums are always positive. Part (b) can be proved substituting $p_{\boldsymbol{Y}} = c^{-1} f$ (the division by $c$ is to make $p_{\boldsymbol{Y}}$ a proper PDF) into the equation for $a$. Similarly Part (c) can be proved by substituting $p_{\boldsymbol{Y}} = \beta^{-d}$ into the equation for $a$. □

Part (a) of implies that we always obtain a positive function when construction an approximation of form $f(\boldsymbol{x}; \boldsymbol{z}_n)$ for any sampled $\boldsymbol{z}_n$ from the data generating process for $\boldsymbol{Z}_n$. Unfortunately there is no guarantee that $f(\boldsymbol{x}; \boldsymbol{z}_n)$ is in $\mathcal{D}(\mathbb{R}^d)$ as the coefficients $a_i$ are not enforced to sum up to 1. However, upon obtaining an approximation of form $f(\boldsymbol{x}; \boldsymbol{z}_n)$, we can simply normalize it in order to obtain $\tilde{f}(\boldsymbol{x}; \boldsymbol{z}_n) = (\sum_{i=1}^n a_i)^{-1} f(\boldsymbol{x}; \boldsymbol{z}_n)$, which is in $\mathcal{D}(\mathbb{R}^d)$. The approximation properties of $\tilde{f}(\boldsymbol{x}; \boldsymbol{z}_n)$ versus that of $f(\boldsymbol{x}; \boldsymbol{z}_n)$ is unknown, however.

Part (b) demonstrates that if we ideally know the form of the approximand $f$, then we can construct an approximation $f(\boldsymbol{x}; \boldsymbol{z}_n)$ that has constant coefficients $a_i = c$. Furthermore, if $\int_{\mathbb{K}} f d\boldsymbol{x} = 1$ then we can set $a_i = 1$, for each $i \in [n]$. This implies that the approximation is in fact in $\mathcal{D}\left(\mathbb{R}^d\right)$ but not necessarily in $\mathcal{D}\left(\mathbb{K}\right)$.

Finally, Part (c) of Corollary 1 can be viewed as an analog to the data generating processes that are suggested by [14]. In the aforementioned paper, the authors only consider the sampling of $\boldsymbol{Z}_n$ from uniform distributions over hypercubes.

## 2.1 Kullback-Leibler Divergence

Define the KL divergence between any two functions that satisfy the necessary integrability assumptions, $f \geq 0$ and $g \geq 0$, over some domain $\mathbb{X}$ as $\text{KL}_{\mathbb{X}}(f, g) = \int_{\mathbb{X}} f \log(f/g) d\boldsymbol{x}$. Further, for every $b > 0$, define the class of lower-bounded square-integrable functions as $\mathcal{S}_b(\mathbb{X}) = \{f \in \mathcal{L}_2(\mathbb{X}) : f \geq b\}$. The following lemma was proved in [33].

**Lemma 3.** *Let $b > 0$ and $\mathbb{X} \subset \mathbb{R}^d$. If $f$ and $g$ are two functions in $\mathcal{S}_b(\mathbb{X})$, then*

$$KL_{\mathbb{X}}(f, g) \leq b^{-1} \|f - g\|_{2,\mathbb{X}}^2.$$

Lemma 3 allows us to bound the KL divergence between two functions by their $\mathcal{L}_2$ distance. Subsequently, we may also use the lemma to provide a bound on the expected KL divergence via the expected $\mathcal{L}_2$ distance, also. The following corollary utilizes Lemma 3 in order to provide a specialized version of Theorem 3 for density approximation.

**Corollary 2.** *If $\mathbb{X} \subset \mathbb{R}^d$ and $\mathbb{K} \supset \mathbb{X}$ are compact, $f \in \mathcal{D}\left(\mathbb{R}^d\right) \cap \mathcal{S}_b(\mathbb{K})$ (for some $b > 0$), and $\phi \in \mathcal{D}\left(\mathbb{R}^d\right)$, then: (a) Uniformly over $\mathbb{X}$, we have*

$$f(\boldsymbol{x}) = \lim_{w_1 \to \infty} \cdots \lim_{w_d \to \infty} \int_{\mathbb{R}^d} g(\boldsymbol{x}; \boldsymbol{w}, \boldsymbol{y}) d\boldsymbol{y}$$

*and*

$$\lim_{w_1 \to \infty} \cdots \lim_{w_d \to \infty} \int_{\mathbb{R}^d} g(\boldsymbol{x}; \boldsymbol{w}, \boldsymbol{y}) \, d\boldsymbol{y} = \lim_{\omega \to \infty} \omega^{-d} \int_{\mathbb{R}^d \times \Omega^d} g(\boldsymbol{x}; \boldsymbol{w}, \boldsymbol{y}) \, d\boldsymbol{y} \, d\boldsymbol{w},$$

*where $\Omega = [0, \omega]$ and $g(\boldsymbol{x}; \boldsymbol{w}, \boldsymbol{y}) = \prod_{j=1}^d w_j \phi(w_j x_j - w_j y_j) f(\boldsymbol{y})$. (b) For any fixed $\omega \in \mathbb{R}_+$, there exists data generating processes for $\boldsymbol{W}$ and $\boldsymbol{Y}$ and a real-valued function $a(\boldsymbol{y})$, such that*

$$\mathbb{E}\left[KL_{\mathbb{X}}\left(f(\boldsymbol{x}; \boldsymbol{Z}_n), \frac{1}{\omega^d} \int_{\mathbb{K} \times \Omega^d} g(\boldsymbol{x}; \boldsymbol{w}, \boldsymbol{y}) \, d\boldsymbol{y} \, d\boldsymbol{w}\right)\right] \to 0, \tag{5}$$

*as $n \to \infty$.*

*Proof.* Part (a) is exactly the same as that of Theorem 3. To prove Part (b), we use Lemma 3 to bound

$$\mathrm{KL}_{\mathbb{X}}\left(f(\boldsymbol{x}; \boldsymbol{Z}_n), \frac{1}{\omega^d} \int_{\mathbb{K} \times \Omega^d} g(\boldsymbol{x}; \boldsymbol{w}, \boldsymbol{y}) \, \mathrm{d}\boldsymbol{y} \mathrm{d}\boldsymbol{w}\right) \tag{6}$$

from above, by $\int_{\mathbb{X}} \left[f(\boldsymbol{x}; \boldsymbol{Z}_n) - \frac{1}{\omega^d} \int_{\mathbb{K} \times \Omega^d} g(\boldsymbol{x}; \boldsymbol{w}, \boldsymbol{y}) \, \mathrm{d}\boldsymbol{y} \mathrm{d}\boldsymbol{w}\right]^2 \mathrm{d}\boldsymbol{x}$. Following from the proof of Theorem 3, we know that the expectation of the aforementioned squared integral has finite expectation for well-chosen data generating processes over $\boldsymbol{W}$ and $\boldsymbol{Y}$, and thus is bounded from above by a term of order $O(n^{-1})$. Now, since the expectation operator preserves inequalities (cf. [4, Sec. 3.2]), we have the fact that the expectation of (6) is also bounded from above by a term of order $O(n^{-1})$ times a constant $b^{-1}$, under the same data generating process. The limit (5) is then obtained by taking $n \to \infty$. □

In statistical estimation and inference, it is often advantageous to work with the KL divergence, as opposed to other loss criteria or objectives. This is due to the close connection between the KL divergence and the popular maximum likelihood criterion (see, e.g. [1, Sec. 7.3]). Corollary 2 justifies the use of maximum likelihood estimation (MLE) for density estimation that is presented in the following section.

## 3 Density Estimation

Let $\boldsymbol{X}_1, \ldots, \boldsymbol{X}_N$ be a random sample of $N \in \mathbb{N}$ observations that is drawn from a data generating process with unknown PDF $f \in \mathcal{D}\left(\mathbb{R}^d\right) \cap \mathcal{S}_b\left(\mathbb{K}\right)$, where $\mathbb{K} \subset \mathbb{R}^d$ is compact. Sample parameters $\boldsymbol{W}_1, \ldots, \boldsymbol{W}_n$ from the uniform distribution on $\Omega^d = [0, \omega]^d$, and sample the parameters $\boldsymbol{Y}_1, \ldots, \boldsymbol{Y}_n$ from a distribution with continuous PDF $p_{\boldsymbol{Y}}$ over $\mathbb{K}$. Select a $\phi \in \mathcal{D}\left(\mathbb{R}^d\right)$ and propose an estimator for $f$ of form (1), which we will denote as

$$f\left(\boldsymbol{x}; \boldsymbol{Z}_n, \boldsymbol{\alpha}\right) = \sum_{i=1}^{n} \alpha_i \prod_{j=1}^{d} W_{ij} \phi\left(W_{ij} x_j - W_{ij} Y_{ij}\right), \tag{7}$$

where $\alpha_i \geq 0$ plays the role of the now unknown $A_i/n$, for each $i \in [n]$ (where $n \in \mathbb{N}$ is constant), and $\boldsymbol{\alpha}^\top = (\alpha_1, \ldots, \alpha_n)$.

Fix the realized samples $\boldsymbol{w}_1, \ldots, \boldsymbol{w}_n$ and $\boldsymbol{y}_1, \ldots, \boldsymbol{y}_n$ (i.e. $\boldsymbol{z}_n$), and suppose that we wish to obtain an estimate of form (7) ($\hat{f}\left(\boldsymbol{x}; \boldsymbol{z}_n\right)$, say), which minimizes the average KL divergence

$$\frac{1}{N} \sum_{k=1}^{N} \mathrm{KL}_{\mathbb{R}^d}\left(f\left(\boldsymbol{X}_k\right), f\left(\boldsymbol{X}_k; \boldsymbol{z}_n, \boldsymbol{\alpha}\right)\right),$$

with respect to $\boldsymbol{\alpha}$ under the unit simplex constraint $\boldsymbol{\alpha} \in \mathbb{S}_n$, where

$$\mathbb{S}_n = \left\{\boldsymbol{s}^\top = (s_1, \ldots, s_n) : s_i \geq 0 \text{ for all } i \in [n], \sum_{i=1}^{n} s_i = 1\right\}.$$

We note that the constraint $\boldsymbol{\alpha} \in \mathbb{S}_n$ is natural since the resulting estimator $\hat{f}\left(\boldsymbol{x}; \boldsymbol{z}_n\right)$ will always be an element of $\mathcal{D}\left(\mathbb{R}^d\right)$.

It is well known that minimizing the average KL divergence is equivalent to MLE (see, e.g. [32, Sec. 1.3]). That is, we can write $\hat{f}(\boldsymbol{x}; \boldsymbol{z}_n)$ as $f(\boldsymbol{x}; \boldsymbol{z}_n, \hat{\boldsymbol{\alpha}}_N)$, where

$$\hat{\boldsymbol{\alpha}}_N = \arg\max_{\boldsymbol{\alpha} \in \mathbb{S}_n} l_N(\boldsymbol{\alpha}; \boldsymbol{z}_n), \tag{8}$$

and $l_N(\boldsymbol{\alpha}; \boldsymbol{z}_n) = N^{-1} \sum_{k=1}^{N} \log f(\boldsymbol{X}_k; \boldsymbol{z}_n, \boldsymbol{\alpha})$ is the log-likelihood function. We will refer to $\hat{f}(\boldsymbol{x}; \boldsymbol{z}_n)$ as the maximum likelihood estimator (MLE) of $f(\boldsymbol{x}; \boldsymbol{z}_n, \boldsymbol{\alpha})$, and $\hat{\boldsymbol{\alpha}}_N$ as the MLE of $\boldsymbol{\alpha}$. Note that we do not refer to an MLE of the estimand $f$, itself. This is because $f$ may not have an exact representation of form (7), for a given fixed $n$. However, as $n$ increases, the best approximation of form (7) will become closer in divergence, on average, to $f$.

Additionally, it is notable that $\hat{\boldsymbol{\alpha}}_n$ always exists since $\mathbb{S}_n$ is a compact set and $l_N(\boldsymbol{\alpha}; \boldsymbol{z}_n)$ is a continuous function. Further observe that $l_N(\boldsymbol{\alpha}; \boldsymbol{z}_n)$ is a concave function by composition, since it is the sum of concave functions (e.g. logarithms) of linear functions of $\boldsymbol{\alpha}$. Thus, since $l_N(\boldsymbol{\alpha}; \boldsymbol{z}_n)$ is also twice differentiable, any stationary point is also a global maximum (i.e. any $\boldsymbol{\alpha}^* \in \mathbb{S}_n$ that satisfies $\nabla l_N(\boldsymbol{\alpha}^*; \boldsymbol{z}_n) = \boldsymbol{0}$, where $\nabla(\cdot)$ and $\boldsymbol{0}$ are the gradient operator and zero vector, respectively).

Unfortunately, the stationary points of the log-likelihood function cannot be obtained in closed form. We therefore require an iterative algorithm that is able to obtains a sequence that converges towards the set of stationary points.

### 3.1 Expectation-Maximization Algorithm

Suppose that we wish to maximize some function $g(\boldsymbol{\theta}) = \log(\sum_{i=1}^{n} \theta_i)$, where $\boldsymbol{\theta}^\top = (\theta_1, \ldots, \theta_n) \in \mathbb{R}_+^n$. By the definitions of [17], $g$ can be minorized at $\boldsymbol{\psi} \in \mathbb{R}_+^n$ by

$$Q(\boldsymbol{\theta}; \boldsymbol{\psi}) = \sum_{i=1}^{n} \tau_i(\boldsymbol{\psi}) \log(\theta_i) - \sum_{i=1}^{n} \tau_i(\boldsymbol{\psi}) \log \tau_i(\boldsymbol{\psi}), \tag{9}$$

where $\tau_i(\boldsymbol{\psi}) = \psi_i / \sum_{j=1}^n \psi_j$. This minorizer is the special case of the Jensen's inequality minorizer (cf. [17, Sec. 4.3]).

Let $\boldsymbol{\alpha}^{(0)} \in \mathbb{S}_n$ be some initial value and $\boldsymbol{\alpha}^{(r)} \in \mathbb{S}_n$ be the $r$th iterate of our EM algorithm. Upon application of (9) on the log-likelihood at the $r$th iteration, we obtain the minorizer

$$Q\left(\boldsymbol{\alpha}; \boldsymbol{\alpha}^{(r-1)}\right) = \sum_{k=1}^N \sum_{i=1}^n \tau_i\left(\boldsymbol{x}_k; \boldsymbol{\alpha}^{(r-1)}\right) \log\left[\alpha_i \prod_{j=1}^d w_{ij}\phi\left(w_{ij}x_{kj} - w_{ij}y_{ij}\right)\right]$$
$$- \sum_{k=1}^N \sum_{i=1}^n \tau_i\left(\boldsymbol{x}_k; \boldsymbol{\alpha}^{(r-1)}\right) \log \tau_i\left(\boldsymbol{x}_k; \boldsymbol{\alpha}^{(r-1)}\right),$$

where $\boldsymbol{x}_k^\top = (x_{k1}, \ldots, x_{kd})$ and

$$\tau_i(\boldsymbol{x}; \boldsymbol{\alpha}) = \frac{\alpha_i \prod_{j=1}^d w_{ij}\phi\left(w_{ij}x_j - w_{ij}y_{ij}\right)}{\sum_{k=1}^n \left[\alpha_k \prod_{j=1}^d w_{kj}\phi\left(w_{kj}x_j - w_{kj}y_{kj}\right)\right]}.$$

We obtain the $r$th iterate of the sequence $\{\boldsymbol{\alpha}^{(r)}\}$ by obtaining a solution vector $\boldsymbol{\alpha}^* \in \mathbb{S}_n$, such that $\nabla Q\left(\boldsymbol{\alpha}^*; \boldsymbol{\alpha}^{(r-1)}\right) = \boldsymbol{0}$. Via a calculus argument, we obtain the solution $\boldsymbol{\alpha}^{*\top} = (\alpha_1^*, \ldots, \alpha_n^*)$ (see, e.g. [17, Eqn. 4.4]), where

$$\alpha_i^* = \frac{1}{N} \sum_{k=1}^N \tau_i\left(\boldsymbol{x}_k; \boldsymbol{\alpha}^{(r-1)}\right). \tag{10}$$

The EM algorithm is defined by setting $\boldsymbol{\alpha}^{(r)} = \boldsymbol{\alpha}^*$ at each iteration $r \in \mathbb{N}$ until some convergence criterion or some limiting number of iterations is met. The final iteration is then declared to be the MLE $\hat{\boldsymbol{\alpha}}_N$. We can check that the assumptions of [25, Thm. 1] are met. We have the following convergence result via the the aforementioned theorem and the properties of log-likelihood function that were previously discussed.

**Proposition 1.** *Let $\boldsymbol{\alpha}^{(r)} \in \mathbb{S}_n$ be defined by (10) and let $\boldsymbol{\alpha}^{(\infty)} = \lim_{r \to \infty} \boldsymbol{\alpha}^{(r)}$ to be the limit point of the sequence $\{\boldsymbol{\alpha}^{(r)}\}$. If the sequence $\{\boldsymbol{\alpha}^{(r)}\}$ is initialized* by some valid $\boldsymbol{\alpha}^{(0)} \in \mathbb{S}_n$, then the limit point $\boldsymbol{\alpha}^{(\infty)}$ is a global maximizer of the realized log-likelihood function $l_N(\boldsymbol{\alpha}; \boldsymbol{z}_n)$.

Proposition 1 is a standard convergence result for EM-type algorithms. The convexity of the objective $l_N(\boldsymbol{\alpha}; \boldsymbol{z}_n)$ and the compactness of $\mathbb{S}_n$ allowed us to establish the global optimality of the EM algorithm, defined by (10).

## 3.2 Consistency of the Maximum Likelihood Estimator

We seek to establish the consistency of the MLE $\hat{\boldsymbol{\alpha}}_N$ as $N \to \infty$, for fixed $n$. The following result can be obtained by checking the assumptions of [29, Thm. 5.14].

**Proposition 2.** *Let $\boldsymbol{\alpha}^0 \in \mathbb{S}_n$ satisfy the equation*

$$\mathbb{E} f\left(\boldsymbol{X}; \boldsymbol{Z}_n, \boldsymbol{\alpha}^0\right) = \sup_{\boldsymbol{\alpha} \in \mathbb{S}_n} \mathbb{E} f\left(\boldsymbol{X}; \boldsymbol{Z}_n, \boldsymbol{\alpha}\right),$$

*and define*

$$\mathbb{S}_n^0 = \left\{\boldsymbol{\alpha} \in \mathbb{S}_n : \mathbb{E} f\left(\boldsymbol{X}; \boldsymbol{Z}_n, \boldsymbol{\alpha}\right) = \mathbb{E} f\left(\boldsymbol{X}; \boldsymbol{Z}_n, \boldsymbol{\alpha}^0\right)\right\}.$$

*If $\hat{\boldsymbol{\alpha}}_N$ is an MLE, as defined by (8), then for every $\epsilon > 0$ and compact set $\mathbb{K} \subset \mathbb{S}_n$, we have*

$$\lim_{N \to \infty} \mathbb{P}\left(\sup_{\boldsymbol{\alpha} \in \mathbb{S}_n^0} \|\hat{\boldsymbol{\alpha}}_N - \boldsymbol{\alpha}\| \geq \epsilon \text{ and } \hat{\boldsymbol{\alpha}}_N \in \mathbb{K}\right) = 0. \tag{11}$$

*Proof.* There are two conditions that need to be checked in order to prove the result. Firstly, we verify that the individual log-densities $f(\boldsymbol{X}; \boldsymbol{Z}_n, \boldsymbol{\alpha})$ are continuous in $\boldsymbol{\alpha}$, for all $\boldsymbol{X}$ and $\boldsymbol{Z}_n$. Secondly, we check that $\sup_{\boldsymbol{\alpha} \in \mathbb{S}_n} \mathbb{E} f(\boldsymbol{X}; \boldsymbol{Z}_n, \boldsymbol{\alpha}) < \infty$. Since $p_{\boldsymbol{W}}$ and $p_{\boldsymbol{Y}}$ is continuous and compactly supported, the expectation $\mathbb{E} f(\boldsymbol{X}; \boldsymbol{Z}_n, \boldsymbol{\alpha})$ exists and is finite for every $\boldsymbol{\alpha}$, due to the fact that both $f$ and $\phi$ are in $\mathcal{D}\left(\mathbb{R}^d\right)$. Since $\mathbb{S}_n$ is compact, the supremum also exists and is finite. This completes the proof. □

The additional statement that $\hat{\boldsymbol{\alpha}}_N \in \mathbb{K}$ for compact $\mathbb{K} \subset \mathbb{S}_n$ in (11) is a technical artifact of [29, Thm. 5.14]. We can simply set $\mathbb{K} = \mathbb{S}_n$ and interpret (11) as if the condition was not present. Proposition 2 then states that the MLE converges in probability towards one of the global maxima of $\mathbb{E}f\left(\boldsymbol{X}; \boldsymbol{Z}_n, \boldsymbol{\alpha}\right)$ as $N \to \infty$, for fixed $n$. This implies that the MLE process provides a pointwise asymptotically unbiased estimator of the best fitting approximation of form (7), with fixed $n$ randomly sampled component density functions with $\boldsymbol{Z}_n$ generated from a data generating process that is characterized by PDFs $p_{\boldsymbol{W}}$ and $p_{\boldsymbol{Y}}$.

It may serve as a minor dissatisfaction that we do not state a consistency result that allows for $n$ to increase as $N$ does. We believe that such a result may be available via the use of method of sieve estimators, such as those that are described by [8] and [9]. Such a result would require a great deal more technical expertise than we currently have available. We therefore defer such explorations to future work.

## 4 Simulation Studies

We perform a set of simulation studies in order to provide empirical evidence that support our theorems from Sections 2 and 3. The first of the simulations (Simulation 1) demonstrates how Corollary 1 can be applied to obtain approximations of known PDFs, and the second simulation (Simulation 2) will demonstrate the use of the density estimation ability of the MLE for randomly sampled MMs.

When considering functional approximation using randomized NNs, it has become convention to utilize the following function as a target (see, e.g. [10]

and [19]):

$$f_0(x) = 0.2e^{-(10x-4)^2} + 0.5e^{-(80x-40)^2} + 0.3e^{-(80x-20)^2},$$

for $x \in \mathbb{X}$, where $\mathbb{X} = [0,1]$. Unfortunately, $f_0$ is neither in $\mathcal{D}(\mathbb{R})$, nor in $\mathcal{D}(\mathbb{X})$. In order to produce a target in $\mathcal{D}(\mathbb{R})$, we normalize each of the exponential components that make up the convex sum of $f_0$. The resulting PDF over $\mathbb{R}$ is $f_1$, and has the form

$$f_1(x) = 0.2\varphi\left(\frac{x-\mu_1}{\sigma_1}\right) + 0.5\varphi\left(\frac{x-\mu_2}{\sigma_2}\right) + 0.3\varphi\left(\frac{x-\mu_3}{\sigma_3}\right),$$

where $\varphi(x) = (2\pi)^{-1/2} e^{-x^2/2}$ is the standard normal PDF, $\mu_1 = 4/10$, $\mu_2 = 40/80$, $\mu_3 = 20/80$, $\sigma_1 = 2^{-1/2}/10$, $\sigma_2 = 2^{-1/2}/80$, and $\sigma_3 = 2^{-1/2}/80$. A plot of $f_1$ is provided in Figure 1. We also take $\phi = \varphi$ in all of our approximation constructions.

### 4.1 Simulation 1

We seek to demonstrate the PDF approximating capacity of random functions of form (1), using the data generating processes and coefficients that are described in Corollary1. Let $n \in \{100, 1000, 10000\}$ and let $\omega \in \{50, 100, 200\}$. For each combination of $n$ and $\omega$ in the aforementioned sets, we firstly (i) construct functions of form (1) (i.e. $f(x; \boldsymbol{z}_n)$) by realizing $\boldsymbol{Z}_n$ using a data generating process that is characterized by the PDFs $p_W = 1/\omega$, and and $p_Y = c^{-1}f_1$, where $c = \int_\mathbb{X} f_1 \mathrm{d}\boldsymbol{x} \approx 1$ (recall that $\mathbb{X} = [0,1]$). We set $a(y) \approx 1$ to comply with Part (b) of Corollary 1. The construction is repeated $R = 100$ times for each combination of $n$ and $\omega$, and we average over the $R$ numerical estimates of KL divergences and squared $\mathcal{L}_2$ distances between the target PDF $f_1$ and each constructed approximation $f(x; \boldsymbol{z}_n)$. That is, we average over $R$ realized

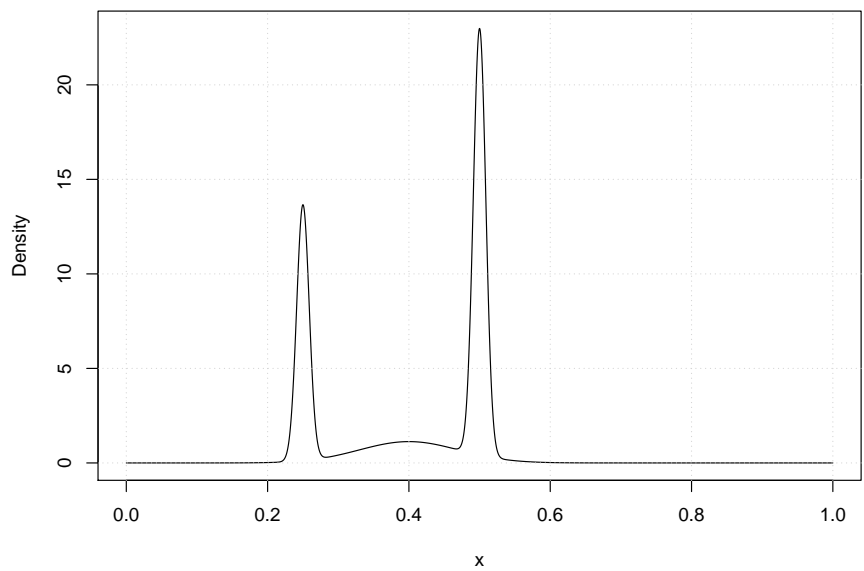

Figure 1: Plot of $f_1$ on the interval $\mathbb{X} = [0, 1]$.

Table 1: Results for Part (i) of Simulation 1.

$\mathbb{A}_{\mathrm{KL}}(n,\omega)$

| $n\backslash\omega$ | 50 | 100 | 200 |
|---|---|---|---|
| 100 | 7.34E-01 | 3.90E-01 | 2.00E-01 |
| 1000 | 7.25E-01 | 3.81E-01 | 1.65E-01 |
| 10000 | 7.24E-01 | 3.77E-01 | 1.62E-01 |

$\mathbb{A}_{\rho}(n,\omega)$

| $n\backslash\omega$ | 50 | 100 | 200 |
|---|---|---|---|
| 100 | 6.44E+00 | 3.84E+00 | 1.83E+00 |
| 1000 | 6.40E+00 | 3.77E+00 | 1.60E+00 |
| 10000 | 6.39E+00 | 3.75E+00 | 1.59E+00 |

numerical estimates of $\mathrm{KL}_{\mathbb{X}}(f_1(x),f(x;\boldsymbol{Z}_n))$ and $\rho^2_{\mathbb{X}}(f_1(x),f(x;\boldsymbol{Z}_n))$. We denote these averages as $\mathbb{A}_{\mathrm{KL}}(n,\omega)$ and $\mathbb{A}_{\rho}(n,\omega)$, respectively, for each $n$ and $\omega$. The numerical estimates are computed via an adaptive Simpson's quadrature method (cf. [24, Ch. 9]). The results of this simulation are reported in Table 1.

Next we (ii) construct functions of form (1) by realizing $\boldsymbol{Z}_n$ using a data generating process that is characterized by the PDFs $p_W = 1/\omega$ and $p_Y = 1$. We then set $a(y) = f_1(y)$, in order to comply with Part (c) of Corollary 1. The construction is repeated $R = 100$ times and we again compute $\mathbb{A}_{\mathrm{KL}}(n,\omega)$ and $\mathbb{A}_2(n,\omega)$, for each combination of $n$ and $\omega$ (in the aforementioned sets). The results of this simulation are reported in Table 2.

## 4.2 Discussions on the Results from Simulation 1

In order to put the numbers from Tables 1 and 2 into context, we compute the KL divergence and $\mathcal{L}_2$ distance between the uniform distribution over $\mathbb{X}$ and the target function $f_1$. These quantities are $\mathrm{KL}_{\mathbb{X}}(f_1, 1) = 1.976$ and $\rho^2_{\mathbb{X}}(f_1, 1) = 10.502$, respectively, and can be used as a benchmark for the computed approximations. In comparison to the benchmarks, we find that both approximations using the data generating processes from Parts (i) and (ii) improve upon the inte-

Table 2: Results for Part (ii) of Simulation 1.

$\mathbb{A}_{\mathrm{KL}}(n,\omega)$

| $n\backslash\omega$ | 50 | 100 | 200 |
|---|---|---|---|
| 100 | 9.03E-01 | 6.44E-01 | 5.51E-01 |
| 1000 | 7.51E-01 | 3.94E-01 | 1.84E-01 |
| 10000 | 7.29E-01 | 3.82E-01 | 1.65E-01 |

$\mathbb{A}_{\rho}(n,\omega)$

| $n\backslash\omega$ | 50 | 100 | 200 |
|---|---|---|---|
| 100 | 7.07E+00 | 5.37E+00 | 4.71E+00 |
| 1000 | 6.52E+00 | 3.91E+00 | 1.88E+00 |
| 10000 | 6.42E+00 | 3.79E+00 | 1.63E+00 |

grated deviations from the uniform distribution. We visualize approximation instances of form (1) for the settings $(n,\omega) \in \{(100, 50), (1000, 100), (10000, 200)\}$, using the protocols from both Parts (i) and (ii), in Figure 2.

From Tables 1 and 2, we observe that both $\mathbb{A}_{\mathrm{KL}}(n,\omega)$ and $\mathbb{A}_{\rho}(n,\omega)$ are decreasing in $n$ and $\omega$. The decrease in $\mathbb{A}_{\rho}(n,\omega)$ due to increases in $\omega$ lends empirical support to Part (a) of Theorem 3, whereupon the approximand $f$ can be expressed as a limit in $\omega$ (i.e. the LHS of Equation (3)). Thus, as $\omega$ gets larger, the LHS of (3) gets closer to the approximand $f$, with respect to the uniform norm.

Decreases in $\mathbb{A}_{\rho}(n,\omega)$ due to increases in $n$ lend empirical support to Part (b) of Theorem 3. This is due to the fact that the limit (4) goes to zero with respect to $n$. Similarly, the decreases in $\mathbb{A}_{\mathrm{KL}}(n,\omega)$ due to $\omega$ and $n$ lend support to Parts (a) and (b) of Corollary 2, respectively, for the same reasons. In Figure 2, we can visually observe the decreases in deviations of the random approximations to the approximand (with respect to increases in $n$ and $\omega$).

In comparing Table 1 to Table 2, we observe that there is an effect due to the choice of the PDF $p_Y$. Setting $p_Y = f_1$ appears to yield better approximations than the uniform sampling scheme characterized by $p_Y = 1$. The choice to set $p_Y = f_1$ is unrealistic in practice, as it requires access to the function $f_1$, which we are trying to approximate for some reason. These reasons may be due to the computational burden of evaluating or simulating from $f_1$ or the lack of a closed for its representation. Thus, to access it in the way that we do is unrealistic in applications of functional approximation. These difficulties arise again in the density estimation computations of Simulation 2, and will be discussed in the sequel.

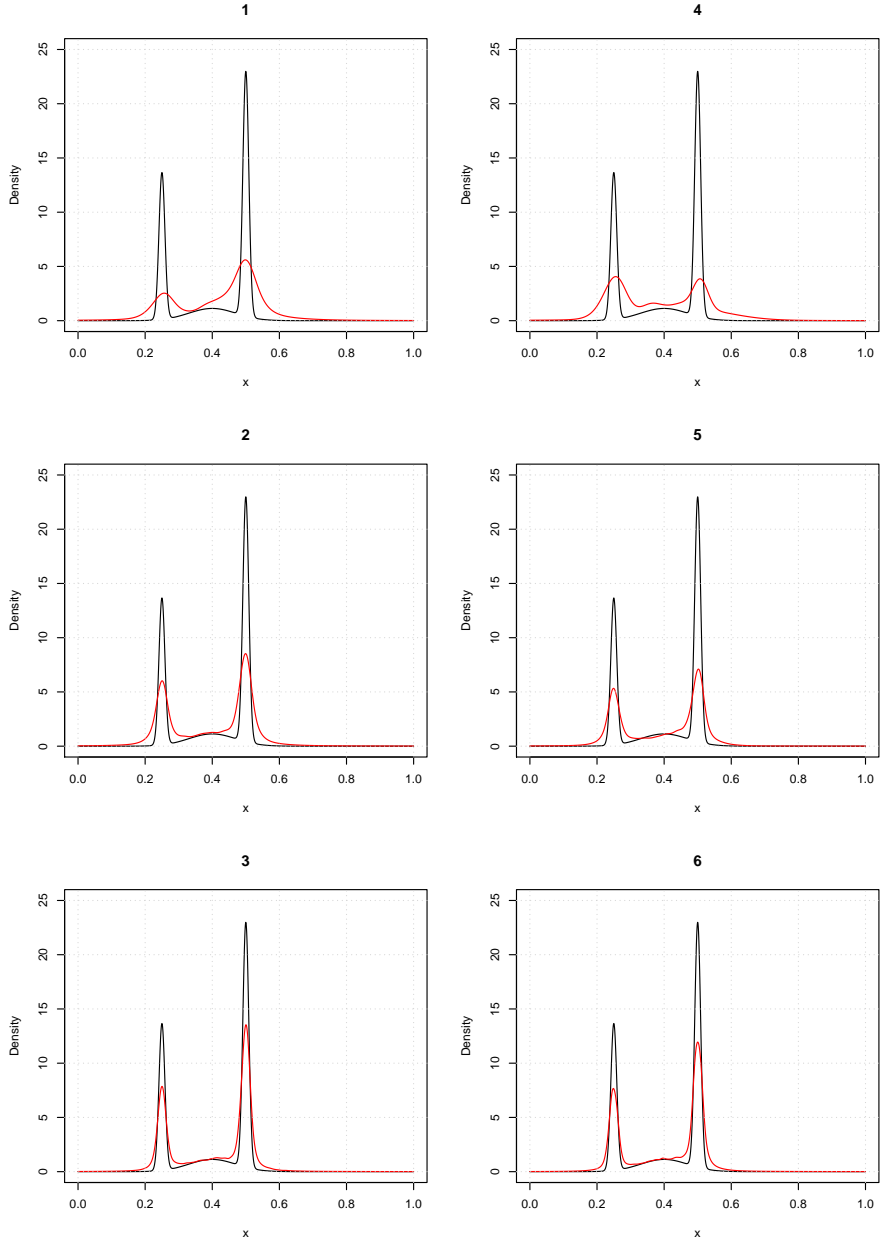

Figure 2: Approximations of $f_1$ (in black) using functions of form (1) (in red) with generating processes described in Parts (i) and (ii) of Simulation 1. Sub-plots 1, 2, and 3 correspond to the generating process described in Part (i), with $(n, \omega)$ set to $(100, 50)$, $(1000, 100)$, and $(10000, 200)$, respectively. Sub-plots 4, 5, and 6 correspond to the generating process described in Part (ii), with $(n, \omega)$ at the same values as those of Sub-plots 1, 2, and 3 (in order).

## 4.3 Simulation 2

We now seek to demonstrate the density estimation capacity of MLEs of random MMs of form (7), fitted using the EM algorithm that is described in Section 3. We set $\omega = 200$ and let $n \in \{100, 500, 1000\}$ and $N \in \{1000, 5000, 10000\}$. For each combination of $n$ and $N$, we realize a sample $x_1, \ldots, x_N$, where each $x_k$ ($k \in [N]$) is a realization of the random variable $X$ that is generated from the distribution that is characterized by the PDF $f_1$. We then obtain a realization $z_n$ of the random variable $Z_n$ in order to construct an estimator of form (7). The components $w_i$ ($i \in [n]$) are realizations of the random variable $W$ with PDF $p_W = 1/\omega$. The components $y_i$ are realizations of the random variable $Y$ with PDF (i) $p_Y = 1$ (i.e. uniformly distributed over $\mathbb{X} = [0, 1]$).

Using the EM algorithm that is defined by Equation (10), we then estimate the parameter vector $\boldsymbol{\alpha}$ by the MLE $\hat{\boldsymbol{\alpha}}_N$ in order to obtain the density estimator $\hat{f}(\boldsymbol{x}; \boldsymbol{z}_n)$. We repeat the process $R = 100$ times for each combination of $n$ and $N$, we compute averages over $R$ realized numerical estimates of $\mathrm{KL}_\mathbb{X}\left(f_1(x), \hat{f}(x; \boldsymbol{Z}_n)\right)$ and $\rho^2_\mathbb{X}\left(f_1(x), \hat{f}(x; \boldsymbol{Z}_n)\right)$. We denote these averages as $\mathbb{A}_{\mathrm{KL}}(n, N)$ and $\mathbb{A}_\rho(n, N)$, respectively, for each $n$ and $N$. Results of the described simulation are presented in Table 3.

As noted in Section 4.2, there are potential differences in achievable approx-imation accuracy levels by way of varying the sampling distribution $p_Y$. Since the target PDF is unknown in any given estimation problem (or else it would be trivialized), the uniform PDF $p_Y = 1$ is a natural choice and somewhat obvious choice.

Table 3: Results for Part (i) of Simulation 2.

| | $\mathbb{A}_{\mathrm{KL}}(n, N)$ | | |
|---|---|---|---|
| $n \backslash N$ | 1000 | 5000 | 10000 |
| 100 | 2.17E-01 | 1.99E-01 | 1.95E-01 |
| 500 | 2.41E-02 | 1.76E-02 | 1.57E-02 |
| 1000 | 1.50E-02 | 7.51E-03 | 6.67E-03 |

| | $\mathbb{A}_{\rho}(n, N)$ | | |
|---|---|---|---|
| $n \backslash N$ | 1000 | 5000 | 10000 |
| 100 | 2.40E+00 | 2.40E+00 | 2.43E+00 |
| 500 | 2.67E-01 | 2.39E-01 | 2.18E-01 |
| 1000 | 1.29E-01 | 6.86E-02 | 7.13E-02 |

Alternatively, and less obviously, is the choice to realize $y_i$ via a random variable with distribution function equal to the empirical distribution function constructed from the realized sample $x_1, \ldots, x_N$ (i.e. $F_N(x) = N^{-1} \sum_{i=1}^{N} [x_i \leq x]$), where $[\varsigma]$ is the Iverson bracket, which takes value 1 if the statement $\varsigma$ is true and 0, otherwise. By the Glivenko-Cantelli theorem, it is known that $F_N$ uniformly converges towards the distribution function of $X$, almost surely. Thus, for sufficiently large $N$, the characterization of $Y$ as a process with distribution $F_N$ closely approximates the characterization of $Y$ as a process with $p_Y$ equal to the target PDF. This process is closely related to the well-known bootstrap method for statistical inference (see, e.g. [7]). We repeat the process described in Part (i), with the uniform sampling of $y_i$ replaced by a sampling from a process with distribution $F_N$. The resulting random MMs share some similarity in concept to the proportion and bandwidth-varying versions of the maximum likelihood kernel density estimators that were proposed by [15]. We refer to the simulation using this new generating process as Part (ii), and we report on its results in Table 4.

Table 4: Results for Part (i) of Simulation 2.

| $\mathbb{A}_{\text{KL}}(n, N)$ | | | |
|---|---|---|---|
| $n\backslash N$ | 1000 | 5000 | 10000 |
| 100 | 1.75E-02 | 1.11E-02 | 1.09E-02 |
| 500 | 1.24E-02 | 4.73E-03 | 3.85E-03 |
| 1000 | 1.22E-02 | 4.30E-03 | 3.54E-03 |

| $\mathbb{A}_{\rho}(n, N)$ | | | |
|---|---|---|---|
| $n\backslash N$ | 1000 | 5000 | 10000 |
| 100 | 8.24E-02 | 5.01E-02 | 4.55E-02 |
| 500 | 6.25E-02 | 2.06E-02 | 1.57E-02 |
| 1000 | 7.21E-02 | 1.71E-02 | 1.27E-02 |

## 4.4 Discussions on the Results from Simulation 2

Upon inspection of Tables 3 and 4, we observe that all approximations have small deviations towards the approximand $f_1$, when comparing with the uniform benchmark results that were reported in Section 4.2. Further, when compared to the results that are presented in Tables 1 and 2, the random density estimators yield approximations that can have KL divergences and $\mathcal{L}_2$ distances that are multiple orders of magnitude smaller than those achieved by the random approximations of Section 4.1. We visualize approximation instances of form (7), for the settings $(n, N) \in \{(100, 1000), (500, 5000), (1000, 10000)\}$, using the protocols from both Parts (i) and (ii), in Figure 2.

Both Tables 3 and 4 indicate that $\mathbb{A}_{\text{KL}}(n, N)$ and $\mathbb{A}_{\rho}(n, N)$ are decreasing in $N$, which lends empirical support to the consistency result of Proposition 2, which indicates that the estimator $\hat{f}(\boldsymbol{x}; \boldsymbol{z}_n)$ provides an increasingly accurate pointwise approximation to the optimal random MM $\mathbb{E}f(\boldsymbol{X}; \boldsymbol{Z}_n, \boldsymbol{\alpha}_0)$, as $N$ increases, for each fixed number of components $n$. We observe that $\mathbb{A}_{\text{KL}}(n, N)$ and $\mathbb{A}_\rho(n, N)$ are decreasing in $n$ in every column of Tables 3 and 4, except for the $N = 1000$ column of the $\mathbb{A}_\rho(n, N)$ results in Table 4. This would again lend support to Theorem 3 and Corollary 2, which both suggest that accuracy, as measured by both the KL divergence and the $\mathcal{L}_2$ distance, can be improved by increasing $n$.

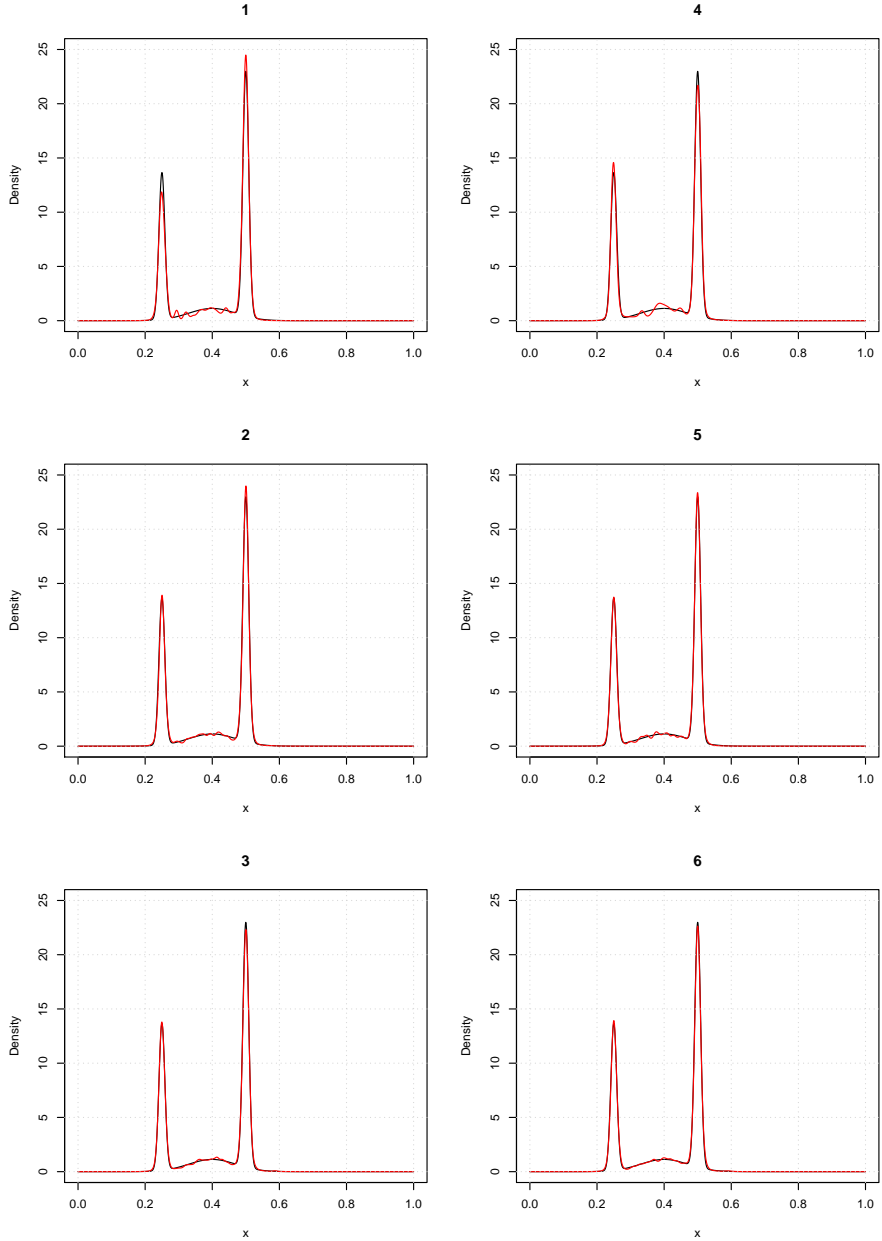

Figure 3: Estimations of $f_1$ (in black) using MMs of form (7) (in red) with generating processes described in Parts (i) and (ii) of Simulation 2. Sub-plots 1, 2, and 3 correspond to the generating process described in Part (i), with $(n, N)$ set to $(100, 1000)$, $(500, 5000)$, and $(1000, 10000)$, respectively. Sub-plots 4, 5, and 6 correspond to the generating process described in Part (ii), with $(n, N)$ at the same values as those of Sub-plots 1, 2, and 3 (in order).

The case of the $N = 1000$ column of the $\mathbb{A}_\rho(n, N)$ results in Table 4 can perhaps be explained by the overfitting phenomenon (see, e.g. [11]). Here, there are too few observations (i.e. $N = 1000$) that are being used to construct models with numbers of parameter elements that are too large. Thus, idiosyncrasies of the data are being modeled and hence the obtained estimators are not generally representative of the target PDF $f_1$. We therefore recommend that a sensible number $N$ be used when constructing random MMs of form (7).

## 5 Conclusions

Although popular, deep NNs can often be difficult to implement due to the complexity of the structures requiring complicated training processes. Because of these complications, the research area of randomized shallow NNs have become increasingly popular.

A popular and theoretically well-founded framework for randomized NNs is the RFVL construction of [14]. In this article, we proved a theorem that removes a limitation of the radial-basis form of the RFVL framework, which restricted its only being applicable for approximation of functions with supports in the unit hypercube. We also specialized the RFVL framework for application to the problem of approximating PDFs. Via a theorem of [33], we demonstrated that the usual $\mathcal{L}_2$ convergence of the RFVL framework can be replaced by a convergence in KL divergence, which is more natural in the problem domain of PDF approximation.

Extending upon the random functional approximation results, we also demonstrated that one can construct random density estimators via mixture modeling and MLE. We derived an EM algorithm for computing the MLE of our random MMs and demonstrated that the EM algorithm is globally convergent towards a global maximum of the log-likelihood function. A consistency result for the MLE is also proved.

A set of simulation studies was presented. The first study provided empirical evidence towards the random approximation theory that we had proved. The second study provided empirical evidence towards the estimation capability of the random MMs.

We note that we had consciously and purposefully left out any comparison between our methodologies and established methods for functional approximation or density estimation. This choice was made due to the novelty of our approach, and its immaturity. We believe that the theoretical results that we have presented are the prime elements of interest and wish to concentrate the attention of the audience on these aspects of our article. We believe that the inclusion of an abundant battery of comparisons at this stage of methodological development would be premature and superfluous.

To conclude the article, we note that there are a number of identifiable directions that require further investigation. The first of these directions is to pursue potential asymptotics that allow for increasing values of $n$, as a function of $N$. This can be achieved, as earlier mentioned, via the method of sieves that is described in [8] and [9]. Another potential direction is to investigate alternative density estimation process to the MLE approach that was discussed in Section 3. Although successful, as was discussed in Section (4.4), the MLE can be computationally burdensome for very large $n$ and $N$, simultaneously. One method to alleviate this burden is to adapt a greedy estimation methods as was considered in [18] and [34]. Greedy procedures for random NNs have been applied to successful effect in the incremental RVFLs of [19] and the SCNs of [31]. The exploration of greedy estimation procedures for random MMs is therefore an interesting and potentially fruitful direction for future work.